\documentclass[aps,english,prd,floatfix,amsmath,eqsecnum, twocolumn, longbibliography]{revtex4-1}

\usepackage{amsmath}
\usepackage{amssymb}
\usepackage{amsthm}
\usepackage{amsfonts}

\usepackage{graphicx,color,framed}
\usepackage{hyperref}
\usepackage{times}
\usepackage{enumerate}
\usepackage{lipsum}
\usepackage{slashed}
\usepackage{url}
\usepackage{comment}
\usepackage{bbm}

\hypersetup{
    colorlinks=true, 
    linktoc=all,     
    linkcolor=blue,  
}

\newcommand{\be}{\begin{equation} }
\newcommand{\ee}{\end{equation} }
\newcommand{\ba}{\begin{eqnarray} }
\newcommand{\ea}{\end{eqnarray} }

\newcommand{\bpm}{\begin{pmatrix}}
\newcommand{\epm}{\end{pmatrix}}
\newcommand{\bmm}{\begin{matrix}}
\newcommand{\emm}{\end{matrix}}

\begin{document}

\title{Physical proof of the topological entanglement entropy inequality}
\author{Michael Levin}

\affiliation{Kadanoff Center for Theoretical Physics, University of Chicago, Chicago, Illinois 60637,  USA}



\begin{abstract}
Recently it was shown that the topological entanglement entropy (TEE) of a two-dimensional gapped ground state obeys the universal inequality $\gamma \geq \log \mathcal{D}$, where $\gamma$ is the TEE and $\mathcal{D}$ is the total quantum dimension of all anyon excitations, $\mathcal{D} = \sqrt{\sum_a d_a^2}$. Here we present an alternative, more direct proof of this inequality. Our proof uses only the strong subadditivity property of the von Neumann entropy together with a few physical assumptions about the ground state density operator. Our derivation naturally generalizes to a variety of systems, including spatially inhomogeneous systems with defects and boundaries, higher dimensional systems, and mixed states.
\end{abstract}

\maketitle

\section{Introduction}

Although we usually think of anyons as \emph{excitations} of two-dimensional (2D) gapped many-body systems, general arguments suggest that their topological properties -- e.g.~fusion rules and braiding statistics -- must be encoded directly in the many-body ground state. Understanding how to extract this anyon data systematically is a challenging problem in the theory of topological phases.

\begin{figure}[tb]
   \centering
   \includegraphics[width=0.5\columnwidth]{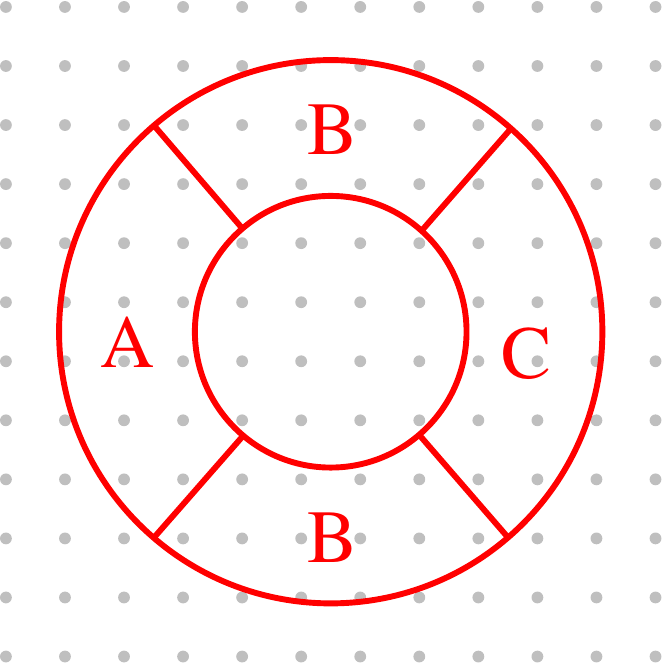}
   \caption{Geometry used to compute the topological entanglement entropy in Eq.~(\ref{eq:TEEdef}).}
   \label{fig:TEE_schematic}
\end{figure}

The topological entanglement entropy (TEE)~\cite{kitaevpreskill2006, levinwen2006} can be regarded as a first step towards solving this problem, as it provides an explicit way to extract one piece of anyon data from a ground state. Let $\rho$ be the density operator for a gapped ground state defined on an infinite lattice. To define the TEE associated with $\rho$, let $A, B, C$ be non-overlapping subsets of the lattice arranged in an annulus geometry as in Fig.~\ref{fig:TEE_schematic}, and consider the conditional mutual information
\begin{equation*}
I(A:C|B)_\rho \equiv S(\rho_{AB}) + S(\rho_{BC}) - S(\rho_B) - S(\rho_{ABC}).
\end{equation*}
Here $\rho_R$ denotes the reduced density operator on region $R$ and $S(\rho_R) = -\text{Tr}( \rho_R \log \rho_R)$ is the von Neumann entropy of $\rho_R$. The topological entanglement entropy $\gamma$ is then defined by\footnote{Here, we use the definition of the TEE given in Ref.~\onlinecite{levinwen2006}. A different definition was given in Ref.~\onlinecite{kitaevpreskill2006}.}
\begin{equation}
\gamma = \frac{1}{2} I(A:C|B)_\rho
\label{eq:TEEdef}
\end{equation}
in the limit of large $A, B, C$.

The original conjecture of Refs.~\onlinecite{kitaevpreskill2006, levinwen2006} was that for any gapped ground state $\rho$, the TEE takes the value $\gamma = \log \mathcal{D}$ where $\mathcal{D} = \sqrt{\sum_a d_a^2}$ is the total quantum dimension of the anyon theory associated with $\rho$, and $d_a$ is the quantum dimension of anyon $a$. This conjecture was supported by field theoretical arguments~\cite{kitaevpreskill2006,dong2008topological}, exact computation of $\gamma$ for a large class of exactly solvable models~\cite{hamma2005bipartite, levinwen2006}, as well as numerical calculations~\cite{Haque2007, furukawa2007}. 

However, it was soon realized that this conjecture is too strong: Refs.~\onlinecite{BravyiUnpublished_2008, Zou_2016, Williamson_2019, Cano_2015, Fliss_2017, Santos_2018, Stephen_2019, Kato_2020} observed that there exist special gapped many-body systems whose TEE $\gamma \neq \log \mathcal{D}$ for arbitrarily large regions $A, B, C$.\footnote{Notably, all known counterexamples are \emph{fine-tuned} in the sense that arbitrarily small perturbations of the parent Hamiltonian restore the equality $\gamma = \log \mathcal D$ in the limit of large $A, B, C$.} These counterexamples showed that the conjecture does not hold universally; more generally, they raised the question of whether there is \emph{any} universal relationship between the TEE and the underlying anyon theory. 

This question was recently answered in the affirmative by Ref.~\onlinecite{teelowerbound}, which proved that there is a universal \emph{inequality} relating $\gamma$ and $\log \mathcal D$:
\begin{equation}
\gamma \geq \log \mathcal{D}.
\label{eq:main_result}
\end{equation}
More specifically, Ref.~\onlinecite{teelowerbound} established the inequality (\ref{eq:main_result}) for any state that can be transformed, via a constant depth circuit, into the ground state of an exactly solvable quantum double~\cite{KITAEV20032} or string-net model~\cite{Levinwen2005, hong2009symmetrization, Hahn2020, LinLevinBurnell2021} or into a state obeying the entanglement bootstrap axioms~\cite{Shi2020}. 

Eq.~(\ref{eq:main_result}) is significant because it establishes a solid theoretical foundation for the TEE similar to that of conventional symmetry-breaking order parameters. To see the analogy, recall that an order parameter with a nonzero expectation value implies that the corresponding symmetry is broken, but a vanishing order parameter does not necessarily imply that the symmetry is \emph{unbroken}. Hence, strictly speaking, order parameters only provide an upper bound to the size of the unbroken symmetry group. Likewise, according to (\ref{eq:main_result}), the TEE provides an upper bound to the total quantum dimension $\mathcal{D}$ of the anyon excitations.

In this paper, we present another proof of (\ref{eq:main_result}) which is complementary to the one given in Ref.~\onlinecite{teelowerbound}. Our proof relies on a small set of physically motivated assumptions about the ground state density operator $\rho$. These assumptions are reasonable in the sense that they follow from the standard theoretical framework for anyons. We show that these assumptions directly imply the inequality (\ref{eq:main_result}) as well as its generalization to the case where $\rho$ contains an anyon $a$ in the center of the annulus:  
\begin{equation}
\gamma \geq \log(\mathcal{D}/d_{a}).
\label{eq:main_result_v2}
\end{equation}

At a high level, the difference between our proof and the one presented in Ref.~\onlinecite{teelowerbound} is that we use a low energy effective theory approach, which relies only on properties of $\rho$ at long length scales, while Ref.~\onlinecite{teelowerbound} uses a microscopic approach, which relies on properties of special zero correlation length ``reference states'' like string-net ground states~\cite{Levinwen2005}. The advantage of our effective theory approach is that it is more direct and does not require any assumptions about the generality of reference states. In addition, our argument can be applied to a variety of systems, including spatially inhomogeneous systems with defects and boundaries, higher dimensional systems, and mixed states. 

This paper is organized as follows. First, in Sec.~\ref{sec:abelian_case} we present our proof in the special case of Abelian states. Then in Sec.~\ref{sec:nonabelian_case} we present the proof for general states. We discuss extensions and generalizations to other systems in Sec.~\ref{sec:discuss}.

\section{Abelian case}
\label{sec:abelian_case}
As a warm-up, we first present a proof of (\ref{eq:main_result}) for the special case where all anyon excitations are Abelian. 

\begin{figure}[tb]
   \centering
   \includegraphics[width=0.4\columnwidth]{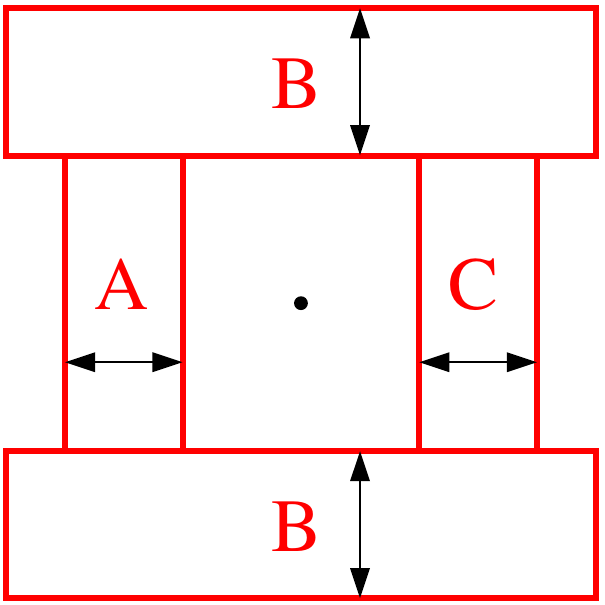} 
   \caption{The class of annuli that we consider: $A, B, C$ are rectangular regions and $ABC$ encloses the origin (black dot). The ``size'' of $ABC$ is the minimum of the widths of regions $A, B, C$ (black lines with arrows) and the distance from the origin to the annulus. }
   \label{fig:rect_annulus}
\end{figure}

\subsection{Setup}
We begin with a few comments about our setup and terminology. These comments apply to both the Abelian case discussed here, and the general case discussed in Sec.~\ref{sec:nonabelian_case}. 

First, regarding the kinds of many-body systems we will consider: throughout the paper, we will focus on two-dimensional \emph{bosonic} lattice models, though our arguments are also applicable to the fermionic case (see Sec.~\ref{sec:discuss}). Also, we will assume that the Hilbert space associated with each lattice site is \emph{finite-dimensional}. This assumption ensures that reduced density operators and von Neumann entropies of subregions are always well-defined and finite. 

Another comment is that when we discuss annular regions $ABC$, we will only consider annuli that enclose a particular fixed point in space -- namely the origin. The reason for this restriction is that we want our setup to be applicable to inhomogeneous systems where the structure of the anyon excitations may depend on the spatial position.

A final point is that, for technical reasons, we will restrict to a particular geometry for our annuli $ABC$: we will assume that $A, B, C$ are \emph{rectangular} regions (or, in the case of $B$, unions of rectangular regions), arranged as in Fig.~\ref{fig:rect_annulus}. This geometry is convenient because it allows us to easily and precisely define the ``size'' of an annulus. Specifically, we will say that an annulus $ABC$ is of ``size $\ell$ or larger'' if the widths of regions $A, B, C$ are at least $\ell$, and the shortest distance from the origin to the annulus is also at least $\ell$.

\subsection{Assumptions}
\label{sec:abel_assump}
We will prove our inequality for any density operator $\rho$ obeying the following assumptions. We will argue below that these assumptions are physically reasonable in the sense that they follow from the the standard theoretical framework for (Abelian) anyons.

Our first assumption is that the density operator $\rho$ can be extended to a \emph{family} of density operators $\{\rho^{(a)} : a \in \mathcal{A}\}$, labeled by elements of a finite Abelian group $\mathcal{A}$, with $\rho = \rho^{(a_0)}$ for some $a_0 \in \mathcal{A}$. Here the physical interpretation of $\mathcal{A}$ is that it describes the set of distinct anyon excitations near the origin, with the group multiplication law in $\mathcal{A}$ corresponding to the fusion product of anyons.

\begin{figure}[tb]
   \centering
   \includegraphics[width=0.92\columnwidth]{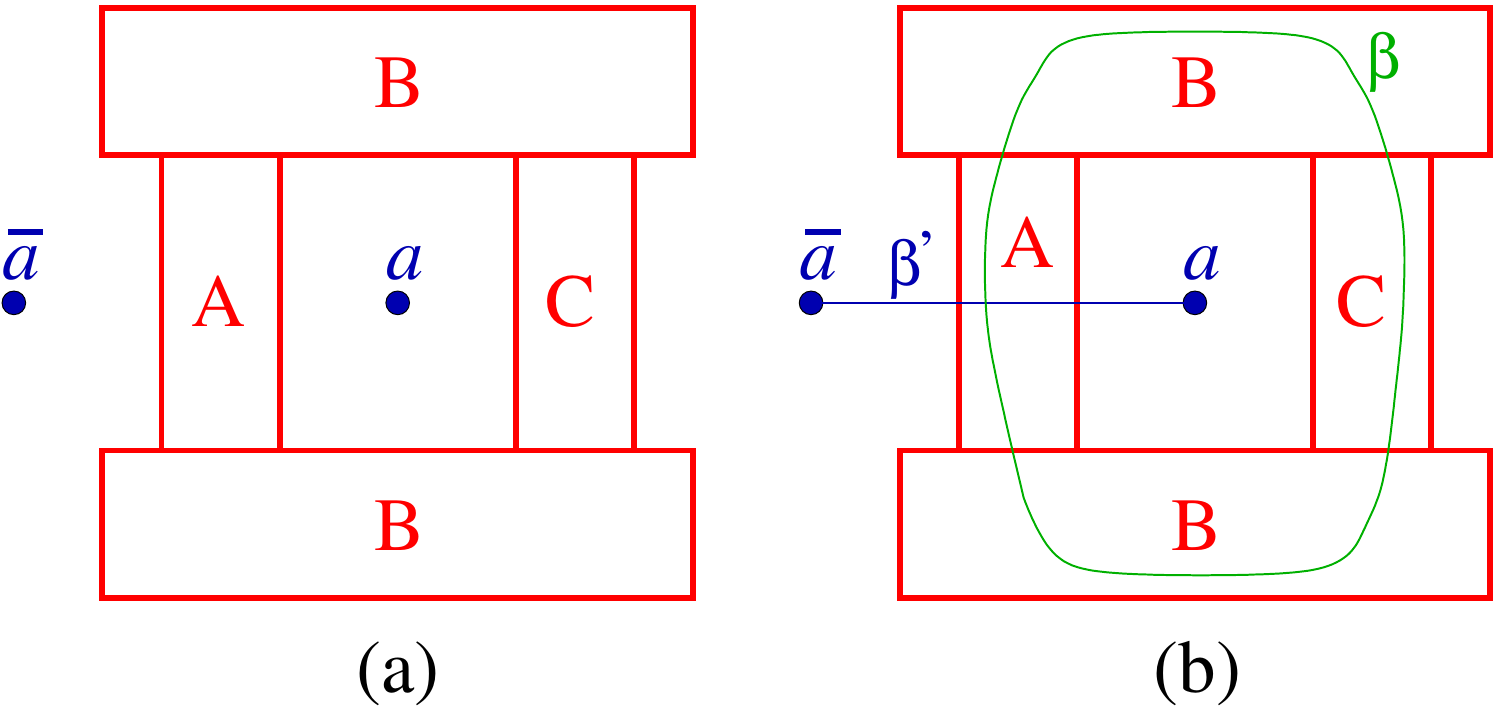} 
   \caption{(a) Physical picture for $\rho^{(a)}$ as a state with an anyon $a$ near the origin and its antiparticle $\bar{a}$ at infinity. (b) Closed and open string operators that can be used to derive global distinguishability and local indistinguishability of $\{\rho^{(a)}\}$ (properties \ref{distabel}, \ref{indistabel} below). }
   \label{fig:rho_a}
\end{figure}

\begin{figure}[t]
   \centering
   \includegraphics[width=0.5\columnwidth]{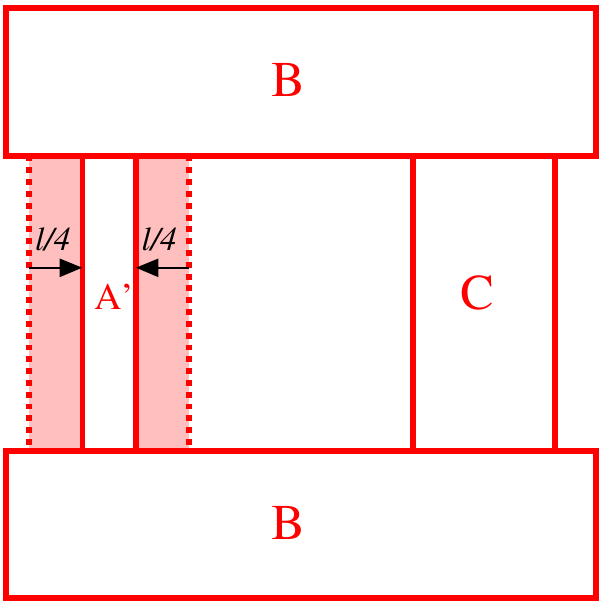} 
   \caption{``Thinning procedure'' for constructing $A'$ from $A$: $A'$ is obtained by (symmetrically) removing the two pink rectangles from $A$, each of width $\ell/4$.}
   \label{fig:AAprime}
\end{figure}

Our second assumption is that the above family of density operators $\{\rho^{(a)}: a \in \mathcal{A}\}$ satisfies the following properties for any annulus $ABC$ of size $\ell$ or larger (where $\ell$ is some fixed length scale associated with $\rho$):
\begin{enumerate}
\item{{\bf Global distinguishability}: $\rho^{(a)}_{ABC}$ and $\rho^{(b)}_{ABC}$ are orthogonal for $a \neq b$.\footnote{Density operators $\rho, \sigma$ are orthogonal if $\text{Tr}(\rho \sigma) = 0$, or equivalently $\text{Tr}(|\rho -\sigma|) = 2$.}} \label{distabel}
\item{{\bf Local indistinguishability}: $\rho^{(a)}_{AB} = \rho^{(b)}_{AB}$ and $\rho^{(a)}_{BC} = \rho^{(b)}_{BC}$ for all $a, b$.} \label{indistabel}
\item{{\bf Fusion}: Let $A' \subset A$ be a thinner version of $A$ with a width reduced by $\ell/2$ as in Fig.~\ref{fig:AAprime}. There exists a set of unitary operators $\{W_s : s \in \mathcal{A}\}$ supported in $A$ such that 
\begin{align}
(W_s \rho^{(a)} W_s^\dagger)_{A'BC} = \rho^{(s \times a)}_{A'BC}
\label{eq:fusionassump}
\end{align}
where $s \times a$ denotes the product of $s, a$ in $\mathcal{A}$.} \label{fusionabel}
\end{enumerate}

To see why these are reasonable assumptions for Abelian states, note that if $\rho$ supports anyon excitations $a \in \mathcal{A}$, we should be able to construct a corresponding family of density operators $\{\rho^{(a)}\}$ describing states with an anyon $a$ near the origin and antiparticle $\bar{a}$ at infinity (Fig.~\ref{fig:rho_a}a). Furthermore, the usual structure of Abelian anyons implies that these $\{\rho^{(a)}\}$ should obey the above properties. In particular, all of the properties follow immediately if we assume the existence of flexible string operators $W_a(\beta)$ that are supported within a finite distance of an open or closed path $\beta$ and that move or create anyon type $a$. For example, property \ref{distabel} says that the different anyon states are \emph{distinguishable} within $ABC$. This distinguishability is guaranteed from the existence of string operators: to see this, consider the closed string operator $W_s(\beta)$ where $\beta$ is the closed path shown in Fig.~\ref{fig:rho_a}b. As long as the annulus is sufficiently thick, then $W_s(\beta)$ is supported entirely within the annulus. Since the $W_s(\beta)$ operators take different eigenvalues in the different anyon states $\rho^{(a)}$ (these eigenvalues define the mutual braiding statistics between $s, a$), it follows that $\rho^{(a)}_{ABC}$ and $\rho^{(b)}_{ABC}$ must be orthogonal for $a \neq b$. 

Likewise, property \ref{indistabel} can be interpreted as the statement that different anyon states are indistinguishable within the subregions $AB$, $BC$. This indistinguishability follows from the existence of unitary open string operators $W_s$ that create pairs of anyons at their endpoints. Using such operators, we can write $\rho^{(a)} = W_s(\beta') \rho W_s^\dagger(\beta')$ where $\beta'$ is the open path shown in Fig.~\ref{fig:rho_a}b and where $s = a \times a_0^{-1}$. Crucially, since the string operators are flexible, we can deform the path $\beta'$ so that it avoids region $AB$ or region $BC$ without affecting $W_s(\beta') \rho W_s^\dagger(\beta')$. It then follows that $\rho^{(a)}$ and $\rho$ are indistinguishable on $AB$ and $BC$, implying property \ref{indistabel}. 

\begin{figure}[t]
   \centering
   \includegraphics[width=0.6\columnwidth]{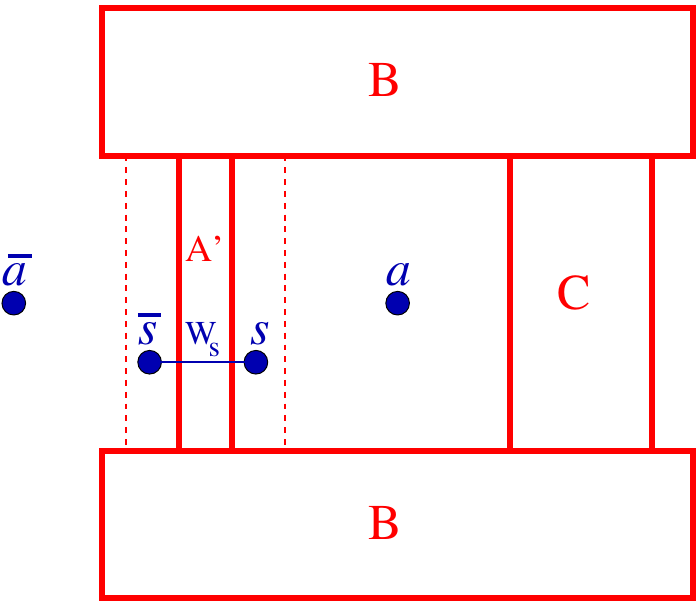} 
   \caption{Physical picture for unitary operator $W_s$ in Eq.~(\ref{eq:fusionassump}): $W_s$ is an open string operator creating anyons $s, \bar{s}$ in region $A \setminus A'$.}
   \label{fig:fusion_assumption}
\end{figure}

As for property \ref{fusionabel}, this is essentially a statement about fusion of anyons. To derive this property from string operators, we can choose $W_s$ to be a (unitary) open string operator that is supported within $A$ and creates a pair of anyons $s, \bar{s}$ in the region $A \setminus A'$ (see Fig.~\ref{fig:fusion_assumption}). Consider the state $W_s \rho^{(a)} W_s^\dagger$. This state contains anyons $s, a$ in the center of the annulus $A'BC$ and anyons $\bar{s}, \bar{a}$ outside of $A'BC$. Therefore, $W_s \rho^{(a)} W_s^\dagger$ can be converted into the state $\rho^{(s \times a)}$ using unitary operators that do not touch $A'BC$. Hence the two states must have the same reduced density operator within $A'BC$, implying (\ref{eq:fusionassump}) above.  

Before concluding this section, we should mention that the above discussion relies on a simplified picture where the anyonic string operators $W_a(\beta)$ are \emph{strictly} supported within a finite distance of the path $\beta$. This picture holds for large classes of states like those that can be transformed into string-net ground states~\cite{Levinwen2005} using a constant depth circuit. However, this picture is not quite correct for generic gapped ground states: in general, the string operators $W_a(\beta)$ are only \emph{approximately} supported within a finite distance of $\beta$, due to the presence of exponentially decaying tails~\cite{hastingswen2005}. Therefore, to treat generic gapped ground states, we need to weaken these assumptions, replacing the exact equalities in properties \ref{distabel}-\ref{fusionabel} with approximate equalities that hold up to some error $\varepsilon$ (in the trace norm) that goes to zero with increasing $\ell$. We believe that our proof can accommodate these extra $\varepsilon$'s, at the cost of only slightly weakening the resulting bound, but we will not discuss this issue here.

\subsection{Statement of (Abelian) claim}
We can now state our claim precisely. Let $\rho$ be any density operator obeying the above assumptions for some Abelian group $\mathcal{A}$ and some length scale $\ell$. Then, for any annulus $ABC$ of size $3\ell/2$ or larger,
\begin{align}
I(A:C|B)_\rho \geq \log |\mathcal{A}|.
\label{eq:abelineq}
\end{align}
(Here $|\mathcal{A}|$ denotes the number of elements of $\mathcal{A}$).

To see the connection between (\ref{eq:abelineq}) and the inequality (\ref{eq:main_result}) that we quoted in the introduction, recall that the total quantum dimension of an Abelian anyon theory is simply the square root of the number of anyons: $\mathcal{D} = \sqrt{|\mathcal{A}|}$. Substituting this expression for $\mathcal{D}$ into (\ref{eq:abelineq}) and using $\gamma = \frac{1}{2} I(A:C|B)_\rho$, gives exactly (\ref{eq:main_result}).

\subsection{Proof}

\subsubsection{Two key inequalities}
We now prove (\ref{eq:abelineq}). The core of the proof consists of two inequalities, which we will derive below. The first inequality gives a lower bound on the conditional mutual information $I(A:C|B)_{\rho^{(a)}}$, \emph{averaged} over different anyon sectors $\rho^{(a)}$. More specifically, this inequality says that for any annulus $ABC$ of size $\ell$ or larger, and any probability distribution $p_a$ on anyon types $a \in \mathcal{A}$, 
\begin{align}
\sum_a p_a I(A:C|B)_{\rho^{(a)}} \geq H(\{p_a\})
\label{eq:paIa}
\end{align}
where $H(\{p_a\}) \equiv - \sum_a p_a \log p_a$ is the Shannon entropy associated with the distribution $p_a$. 

The second inequality relates the conditional mutual information in two different annuli, namely $ABC$ and $A'BC$ where $A' \subset A$ is the subregion mentioned in property \ref{fusionabel} above. This inequality says that for any annulus $ABC$ of size $\ell$ or larger, and for any $s \in \mathcal{A}$,
\begin{align}
I(A:C|B)_{\rho^{(a)}} \geq I(A':C|B)_{\rho^{(s \times a)}}.
\label{eq:Iasa}
\end{align}
We will see that the inequality (\ref{eq:paIa}) is a direct mathematical consequence of properties \ref{distabel}, \ref{indistabel}, together with the strong subadditivity property of the von Neumann entropy~\cite{Lieb1973}. Likewise, we will see that (\ref{eq:Iasa}) is a direct mathematical consequence of property \ref{fusionabel}. We postpone the derivation of these inequalities to Sec.~\ref{sec:derivabelineq} below.

\subsubsection{Main argument}
Once we establish (\ref{eq:paIa}-\ref{eq:Iasa}), the claim (\ref{eq:abelineq}) follows easily: we simply note that
\begin{align}
I(A:C|B)_{\rho^{(a_0)}} &\geq \frac{1}{|\mathcal{A}|} \sum_s I(A':C|B)_{\rho^{(s \times a_0)}} \nonumber \\
&= \frac{1}{|\mathcal{A}|} \sum_a I(A':C|B)_{\rho^{(a)}} \nonumber \\
&\geq \log |\mathcal{A}|
\label{eq:abelproof}
\end{align}
which proves the claim since $\rho^{(a_0)} = \rho$ by assumption. 
Here, the first line in (\ref{eq:abelproof}) is obtained by considering (\ref{eq:Iasa}) with $a = a_0$ and then averaging this inequality over all $s \in \mathcal{A}$. The second line in (\ref{eq:abelproof}) follows from the fact that $s \times a_0$ runs over all anyon types exactly once so we can replace $\rho^{(s \times a_0)} \rightarrow \rho^{(a)}$ without changing the sum. Finally, the last line follows from applying (\ref{eq:paIa}) to the annulus $A'BC$ with $p_a = \frac{1}{|\mathcal{A}|}$ and noting that the Shannon entropy of the uniform distribution is $H (\{p_a\}) = \log |\mathcal{A} |$. (Note that (\ref{eq:paIa}) is applicable to $A'BC$, since we assumed that $ABC$ is of size $3\ell/2$ or larger, which means that the thinner annulus $A'BC$ is of size $\ell$ or larger). 

\subsubsection{Deriving the inequalities (\ref{eq:paIa}) and (\ref{eq:Iasa})}
\label{sec:derivabelineq}
All that remains is to derive the inequalities  (\ref{eq:paIa}-\ref{eq:Iasa}). To derive (\ref{eq:paIa}), consider a probabilistic mixture of the different anyon sectors:
\begin{align}
\lambda = \sum_a p_a \rho^{(a)} 
\end{align}
where the $p_a$ are probabilities with $\sum_a p_a = 1$, and $p_a \geq 0$. We claim that the conditional mutual information in this state is given by
\begin{align}
I(A:C|B)_\lambda = \sum_a p_a I(A:C|B)_{\rho^{(a)}} - H(\{p_a\}).
\label{eq:iabclambda}
\end{align}
Once we show this, we will be finished since the inequality (\ref{eq:paIa}) follows immediately from (\ref{eq:iabclambda}) together with the strong subadditivity property of von Neumann entropy, $I(A:C|B)_\lambda \geq 0$~\cite{Lieb1973}.

To derive (\ref{eq:iabclambda}) recall that 
\begin{align*}
I(A:C|B)_\lambda = S(\lambda_{AB}) + S(\lambda_{BC}) - S(\lambda_B) - S(\lambda_{ABC}). 
\end{align*}
We now compute each of the above von Neumann entropies. We start with $S(\lambda_{ABC})$. Since $\lambda_{ABC} = \sum_a p_a \rho^{(a)}_{ABC}$ and the $\rho^{(a)}_{ABC}$ are orthogonal (by property \ref{distabel}), it follows that
\begin{align}
S(\lambda_{ABC}) =  \sum_a p_a S(\rho^{(a)}_{ABC}) + H(\{p_a\})
\label{eq:sabclambda}
\end{align}
where $H(\{p_a\}) = - \sum_a p_a \log p_a$ is the Shannon entropy associated with the probability distribution $p_a$. (Here, Eq.~(\ref{eq:sabclambda}) is the standard formula for the von Neumann entropy of a mixture of orthogonal states). 
Next, consider $S(\lambda_{AB})$. Using property \ref{indistabel} above, we can write
\begin{align}
S(\lambda_{AB}) &= S \left(\sum_a p_a \rho^{(a)}_{AB} \right) \nonumber \\
&= S \left(\sum_a p_a \rho^{(a_0)}_{AB} \right) \nonumber \\
&= \sum_a p_a  S(\rho^{(a_0)}_{AB}) \nonumber \\
&= \sum_a p_a  S(\rho^{(a)}_{AB})
\label{eq:sablambda}
\end{align}
where the first equality follows from the definition of $\lambda$, the second and fourth equalities follow from property \ref{indistabel}, and the third equality follows from the normalization $\sum_a p_a = 1$. By the same reasoning
\begin{align}
S(\lambda_{BC}) = \sum_a p_a  S(\rho^{(a)}_{BC}), \quad \quad S(\lambda_{B}) = \sum_a p_a  S(\rho^{(a)}_{B}).
\label{eq:sbclambda}
\end{align}
Combining (\ref{eq:sabclambda}), (\ref{eq:sablambda}), (\ref{eq:sbclambda}) gives (\ref{eq:iabclambda}) above, as we wished to show.

Next, we derive the inequality (\ref{eq:Iasa}). To do this, let $W_s$ be the operator from property \ref{fusionabel} and let $A' \subset A$ be the associated subregion of $A$. Then we have
\begin{align}
I(A:C|B)_{\rho^{(a)}} &= I(A:C|B)_{W_s \rho^{(a)} W_s^\dagger} \nonumber \\
&\geq  I(A':C|B)_{W_s \rho^{(a)} W_s^\dagger} \nonumber \\
&=  I(A':C|B)_{\rho^{(s \times a)}}.
\end{align}
Here the first line follows from the unitarity of $W_s$ and the fact that it is supported entirely within $A$: these two facts imply that $W_s$ does not affect the von Neumann entropies in any of the regions $AB, BC, B, ABC$ and hence $W_s$ does not affect the conditional mutual information $I(A:C|B)_{\rho^{(a)}}$. The second line follows from the general fact that the conditional mutual information $I(A:C|B)$ is \emph{non-increasing} under tracing out spins in $A$ or $C$ (a corollary of the strong subadditivity property of the von Neumann entropy~\cite{Lieb1973}). Finally, the third line follows from (\ref{eq:fusionassump}). This establishes (\ref{eq:Iasa}).

\section{General case}
\label{sec:nonabelian_case}

We now present our proof for general states. The argument is similar in spirit to the Abelian case, but with a few new elements, namely the concepts of fusion probabilities $p_{s \times a \rightarrow b}$ and fixed-point probability distributions $p_a^*$. 
  
\subsection{Assumptions}
\label{sec:nonabelian_assump}
As in the Abelian case, we prove our inequality for any density operator $\rho$ obeying certain assumptions. We will argue below that these assumptions are well-justified in that they follow from the standard theoretical framework for anyons.

Our first assumption is the same as in the Abelian case: we assume that $\rho$ can be extended to a finite collection of density operators $\{\rho^{(a)} : a \in \mathcal{A}\}$, with $\rho = \rho^{(a_0)}$ for some $a_0 \in \mathcal{A}$. As before, the physical interpretation of $\mathcal{A}$ is that it describes the set of distinct anyon excitations near the origin. 

However, unlike the Abelian case, we do not assume that $\mathcal{A}$ has the structure of an Abelian group. Instead, we only assume the existence of a collection of real numbers $p_{s \times a \rightarrow b} \in [0,1]$, defined for any $s, a, b \in \mathcal{A}$, with the property that $\sum_b p_{s \times a \rightarrow b} = 1$. These numbers can be interpreted as fusion probabilities: $p_{s \times a \rightarrow b}$ is the probability that two independently created anyons $s$ and $a$ will fuse to $b$~\cite{preskilltqc, Kitaev_2006}. (To connect this with the Abelian case discussed earlier, note that in the Abelian case all fusion probabilities are either $0$ or $1$ and are given by $p_{s \times a \rightarrow b} = \delta_{b, s \times a}$.) 

We also assume that there exists a strictly positive probability distribution $p_a^* > 0$ satisfying 
\begin{align}
\sum_s p_{s \times a \rightarrow b} \cdot p_s^* = p_b^* 
\label{eq:probid}
\end{align}
for all $a, b \in \mathcal{A}$. The physical interpretation of Eq.~(\ref{eq:probid}) is that the probability distribution $p_a^*$ is a ``fixed-point'' under fusion. That is, if we choose anyons $s$ with probabilities $p_s^*$ and then fuse with some fixed anyon $a$, the resulting anyons $b$ will be governed by the same probability distribution $p_b^*$. We note that this fixed-point distribution $p_a^*$  is guaranteed to be unique by the Perron-Frobenius theorem.

The above assumption (\ref{eq:probid}) about the existence of the fixed-point distribution $p_a^*$ can be motivated in two ways. First, one can derive this property directly from the algebraic theory of anyons~\cite{Kitaev_2006}. Indeed, as we explain in Appendix~\ref{sec:probidapp}, the standard anyon formalism not only implies the existence of $p_a^*$, but it also gives an explicit formula for it:
\begin{align}
p_a^* = \frac{d_a^2}{\mathcal{D}^2}
\label{eq:pastarform}
\end{align}
where $d_a$ is the quantum dimension of $a$, and $\mathcal{D} = \sqrt{ \sum_a d_a^2}$ is the total quantum dimension. One can also argue for the existence of $p_a^*$ on more general grounds. In Appendix~\ref{sec:probidapp2}, we show that $p_a^*$ is guaranteed to exist as long as the fusion probabilities $p_{s \times a \rightarrow b}$ obey two (physically reasonable) conditions: (i) for every $a, b$, there exists at least one $s$ such that $p_{s \times a \rightarrow b} > 0$ and (ii) the fusion probabilities are \emph{associative} in the sense that $\sum_u p_{s \times t \rightarrow u} p_{u \times a \rightarrow c} = \sum_b  p_{t \times a \rightarrow b} p_{s \times b \rightarrow c}$.

Our final assumption is analogous to the Abelian case. We assume that the family of density operators $\{\rho^{(a)}: a \in \mathcal{A}\}$ obeys the following properties for any annulus $ABC$ of size $\ell$ or larger (where $\ell$ is some fixed length scale associated with $\rho$):
\begin{enumerate}
\item{{\bf Global distinguishability}: $\rho^{(a)}_{ABC}$ and $\rho^{(b)}_{ABC}$ are orthogonal for $a \neq b$.} \label{distnonabel}
\item{{\bf Local indistinguishability}: $\rho^{(a)}_{AB} = \rho^{(b)}_{AB}$ and $\rho^{(a)}_{BC} = \rho^{(b)}_{BC}$ for all $a, b$.} \label{indistnonabel}
\item{{\bf Fusion}: Let $A' \subset A$ be a thinner version of $A$ with a width reduced by $\ell/2$ as in Fig.~\ref{fig:AAprime}. There exists a set of unitary operators $\{W_s : s \in \mathcal{A}\}$ supported in $A$ such that 
\begin{align}
(W_s \rho^{(a)} W_s^\dagger)_{A'BC} = \sum_{b} p_{s \times a \rightarrow b} \cdot \rho^{(b)}_{A'BC}.
\label{eq:fusionassumpnonabel}
\end{align} 
} \label{fusionnonabel}
\end{enumerate}
Properties \ref{distnonabel} and \ref{indistnonabel} are identical to the Abelian case and they can be motivated in the same way as in Sec.~\ref{sec:abel_assump}. As for property \ref{fusionnonabel}, this is a generalization of the fusion assumption (\ref{eq:fusionassump}) in the Abelian case. As in that case, we can derive this property if we assume the existence of unitary open string operators that create pairs of anyons. To do this, we choose $W_s$ to be an open string operator that is supported within $A$ and creates a pair of anyons $s, \bar{s}$ in the region $A \setminus A'$ (see Fig.~\ref{fig:fusion_assumption}). Consider the state $W_s \rho^{(a)} W_s^\dagger$. This state contains anyons $s, a$ in the center of the annulus $A'BC$ and anyons $\bar{s}, \bar{a}$ outside of $A'BC$. In general $s, a$ can fuse to multiple possible anyons $b$ with probability $p_{s \times a \rightarrow b}$, so we expect that the reduced density operator $(W_s \rho^{(a)} W_s^\dagger)_{A'BC}$ is given by a probabilistic mixture $\sum_b p_{s \times a \rightarrow b} \cdot \rho^{(b)}_{A'BC}$ as in (\ref{eq:fusionassumpnonabel}) above. 

We note that the same caveat applies here as in the Abelian case: although our assumptions hold exactly for many states (e.g.~states that can be connected to string-net ground states~\cite{Levinwen2005} using a constant depth circuit), we only expect them to hold \emph{approximately} for generic gapped ground states. To correctly treat generic gapped ground states, we would need to weaken properties \ref{distnonabel}-\ref{fusionnonabel} by including error terms in each of the equalities that go to zero with increasing $\ell$. We expect that our proof can readily accommodate such error terms, at the cost of slightly weakening the resulting bound, but we will not discuss this issue here.

\subsection{Statement of claim}

We can now state our main claim precisely. Let $\rho$ be any density operator obeying the above assumptions for some set of anyons $\mathcal{A}$ and length scale $\ell$. Then for any annulus $ABC$ of size $(n+3)\ell/2$ or larger (with $n \geq 1$),
\begin{align}
I(A:C|B)_{\rho} \geq \log\left (\frac{1}{p^*_{a_0}}\right) - \frac{K}{\sqrt{n}}
\label{eq:nonabelineq}
\end{align}
where $p^*_a$ is the fixed-point probability distribution from our assumption (\ref{eq:probid}) and $\rho = \rho^{(a_0)}$. Here $K$ is the order one constant
\begin{align} 
K = 1 + \frac{2}{\mathfrak{p}^2} \log\left(\frac{|\mathcal{A}|}{\mathfrak{p}}\right)
\end{align}
where $\mathfrak{p} = \min_a p_a^*$.

To see the connection between this result and the inequalities (\ref{eq:main_result}) and (\ref{eq:main_result_v2}) from the introduction, recall that $p_a^* = \frac{d_a^2}{\mathcal{D}^2}$ (\ref{eq:pastarform}) for any system described by the standard theoretical framework for anyons. Making this identification, (\ref{eq:nonabelineq}) becomes
\begin{align}
I(A:C|B)_{\rho} \geq 2 \log\left (\frac{\mathcal{D}}{d_{a_0}}\right)  - \frac{K}{\sqrt{n}}.
\label{eq:nonabelineqd}
\end{align}
Clearly this inequality is equivalent to (\ref{eq:main_result_v2}) up to a correction $\frac{K}{\sqrt{n}}$ which vanishes in the thermodynamic limit. Likewise, if we set $a_0$ to be the trivial anyon (as would be appropriate for a typical spatially homogeneous ground state) this inequality reduces to (\ref{eq:main_result}) in the thermodynamic limit. 

One comment about our main result (\ref{eq:nonabelineq}): while this inequality is certainly tight in the thermodynamic limit $n \rightarrow \infty$ (for example, it is saturated by string-net ground states~\cite{levinwen2006}) we do not know to what extent it is tight for finite $n$. That is, it is possible that the correction term $K/\sqrt{n}$ on the right hand side can be replaced with a term that goes to zero faster as $n \rightarrow \infty$, or perhaps it can be removed entirely. Indeed, Ref.~\onlinecite{teelowerbound} showed that all states that are connected to string-net ground states by a constant depth circuit -- the simplest class of states that obey our assumptions -- obey (\ref{eq:nonabelineq}) without the $K/\sqrt{n}$ term.

\subsection{Proof}

\subsubsection{Two key inequalities}
As in the Abelian case, our proof of (\ref{eq:nonabelineq}) is based on two inequalities which we will derive below. The first inequality is identical to Eq.~(\ref{eq:paIa}). It says that, for any annulus $ABC$ of size $\ell$ or larger, and any probability distribution $p_a$ on anyon types,
\begin{align}
\sum_a p_a I(A:C|B)_{\rho^{(a)}} \geq H(\{p_a\})
\label{eq:paIarepeat}
\end{align}
where $H(\{p_a\}) = -\sum_a p_a \log p_a$ is the Shannon entropy. 

The second inequality can be thought of as a non-Abelian generalization of Eq.~(\ref{eq:Iasa}). This inequality says that for any annulus $ABC$ of size $\ell$ or larger, any probability distribution $p_a$, and any anyon $s \in \mathcal{A}$,
\begin{align}
\sum_a p_a &I(A:C|B)_{\rho^{(a)}} - H(\{p_a\}) \geq \nonumber \\
&\sum_a p_{a,s} I(A':C|B)_{\rho^{(a)}} -H(\{p_{a,s}\}) 
\label{eq:Iasanonabel}
\end{align}
where $p_{a,s}$ is a probability distribution on $a$ defined by
\begin{align}
p_{a,s} = \sum_b p_b \cdot p_{s \times b \rightarrow a}.
\label{eq:pas}
\end{align}
The physical meaning of $p_{a,s}$ is that it is the probability of obtaining an anyon $a$ if one picks an anyon $b$ using the probability distribution $p_b$, and then fuses it with $s$. We note that (\ref{eq:Iasanonabel}) reduces to (\ref{eq:Iasa}) in the Abelian case, if we consider the probability distribution $p_a = \delta_{ab}$. In this case, $p_{a,s} = \delta_{a, b\times s}$, so $H(\{p_a\}) = H(\{p_{a,s}\}) = 0$.  

Similarly to the Abelian case, we will see that (\ref{eq:paIarepeat}) is a direct consequence of properties \ref{distnonabel}, \ref{indistnonabel} together with strong subadditivity, while (\ref{eq:Iasanonabel}) is a consequence of property \ref{fusionnonabel}. We postpone the derivation of these inequalities to Sec.~\ref{sec:derivnonabelineq} below.

\subsubsection{Main argument}
We now use the two inequalities (\ref{eq:paIarepeat}) and (\ref{eq:Iasanonabel}) to prove our claim (\ref{eq:nonabelineq}). Conceptually, this proof is similar to the Abelian case, but the technical details are more complicated. In particular, one key difference is that we will use the inequalities (\ref{eq:paIarepeat}) and (\ref{eq:Iasanonabel}) \emph{multiple} times, rather than once, as we did in the Abelian case.

The idea is as follows. Given an annulus $ABC$ of size $(n+3)\ell/2$ or larger, we can construct a sequence of nested annuli, $A_0BC \subset A_1BC \subset \cdots \subset A_{n+1}BC = ABC$, where $A_{n+1} \equiv A$, and the other $A_i$'s are defined recursively by $A_i = A_{i+1}'$ with $A_{i+1}'$ given by the thinning procedure shown in Fig.~\ref{fig:AAprime}. By construction, each pair of annuli $A_iBC \subset A_{i+1}BC$ gives an inequality like (\ref{eq:Iasanonabel}), namely
\begin{align}
\sum_a p_a I^{(a)}_{i+1} - H(\{p_a\}) \geq \sum_a p_{a,s} I^{(a)}_i -H(\{p_{a,s}\}) 
\label{eq:Iasanonabeli}
\end{align}
where $I^{(a)}_i \equiv I(A_i:C|B)_{\rho^{(a)}}$. Likewise, applying (\ref{eq:paIarepeat}) to each annulus $A_iBC$, gives the inequality
\begin{align}
\sum_a p_a I^{(a)}_i \geq H(\{p_a\}).
\label{eq:paIai}
\end{align}
We will show that by combining the inequalities (\ref{eq:Iasanonabeli}) and (\ref{eq:paIai}), we can derive a strong inequality for $I^{(b)}_{n+1}$. In particular, we will show below that the inequalities (\ref{eq:Iasanonabeli}) and (\ref{eq:paIai}) imply that
\begin{align}
I^{(b)}_{n+1} \geq  \log \left(\frac{1}{p_b^*}\right) - \frac{K}{\sqrt{n}}.
\label{eq:nonabelineq2}
\end{align}
Since $I^{(b)}_{n+1} = I(A:C|B)_{\rho^{(b)}}$, this is equivalent to our claim (\ref{eq:nonabelineq}). 

We now explain how this works in detail. We begin by writing down a few special cases/corollaries of (\ref{eq:Iasanonabeli}) which we will need in our argument. The first corollary of (\ref{eq:Iasanonabeli}) that we will use is that
\begin{align}
I^{(b)}_{i+1} \geq I^{(b)}_{i}
\label{eq:Iasanonabeli_cor1}
\end{align}
for any $b \in \mathcal{A}$. This inequality is obtained by considering (\ref{eq:Iasanonabeli}) in the special case $p_a = \delta_{ab}$ and $s = 1$. Another corollary of (\ref{eq:Iasanonabeli}) that we will need is that
\begin{align}
I^{(b)}_{i+1} \geq \sum_a p_a^* I^{(a)}_i - \log |\mathcal{A}| 
\label{eq:Iasanonabeli_cor2}
\end{align}
for any $b \in \mathcal{A}$. To derive this result, we average (\ref{eq:Iasanonabeli}) over different choices of $s$ with relative weights $p_s^*$: that is, we multiply both sides of (\ref{eq:Iasanonabeli}) by $p_s^*$ and sum over $s \in \mathcal{A}$. This gives
\begin{align}
\sum_a p_a I^{(a)}_{i+1} - H(\{p_a\}) \geq \sum_a p_a^* I^{(a)}_{i} - \sum_s p_s^* H(\{p_{a,s}\}) 
\label{eq:Iasanonabeli2}
\end{align}
where we have used the identity $\sum_s p_s^* p_{a,s} = p_a^*$ to simplify the first term on the right hand side. Next, we set $p_a = \delta_{ab}$, so that (\ref{eq:Iasanonabeli2}) simplifies to
\begin{align*}
I^{(b)}_{i+1} \geq \sum_a p_a^* I^{(a)}_i - \sum_s p_s^* H(\{p_{s \times b \rightarrow a}\}).
\end{align*}
Eq.~(\ref{eq:Iasanonabeli_cor2}) now follows immediately, since the Shannon entropy for any probability distribution, is bounded above by the logarithm of the number of possible outcomes: $H(\{p_{s \times b \rightarrow a}\}) \leq \log |\mathcal{A} |$.
 
Our final corollary of (\ref{eq:Iasanonabeli}) is that, for any pair of anyons $b, c \in \mathcal{A}$ and any $\epsilon$ with $|\epsilon| \leq \mathfrak{p}/2$,
\begin{align}
\sum_a p_a^* (I^{(a)}_{i+1} -  I^{(a)}_i) &\geq \epsilon \mathfrak{p} \left[I^{(c)}_{i} - I^{(b)}_i + \log\left( \frac{p_{c}^*}{p_b^*}\right) \right] - 2 \epsilon^2
\label{eq:Iasanonabeli_cor3}
\end{align} 
where $\mathfrak{p} \equiv \min_a p_a^*$.
We derive this corollary in Appendix~\ref{sec:nonabelcorapp} by substituting the probability distribution of the form $p_a = p_a^* + \epsilon \delta_{a b} - \epsilon \delta_{a c}$ into (\ref{eq:Iasanonabeli2}), and then working out the first order Taylor series expansion for $H(\{p_a\})$ and $H(\{p_{a,s}\})$ (in terms of $\epsilon$).

With these corollaries, we are now ready to complete the proof. The first step is to multiply (\ref{eq:Iasanonabeli_cor3}) by $p_c^*$ and sum over all $c$. The result can be written in the abbreviated form
\begin{align}
\sum_a p_a^* (I^{(a)}_{i+1} -  I^{(a)}_i) \geq \epsilon \mathfrak{p}  \delta_i^{(b)} - 2 \epsilon^2
\label{eq:Iasanonabeli_cor4}
\end{align} 
where the quantity $\delta_i^{(b)}$ is defined by
\begin{align}
\delta_i^{(b)} \equiv \sum_a p_a^* \left[I_i^{(a)} -I_i^{(b)} + \log\left( \frac{p_{a}^*}{p_b^*}\right)\right]. 
\end{align}
Next, we sum (\ref{eq:Iasanonabeli_cor4}) over $i = 0,...,n-1$ and use the fact that $I^{(a)}_0 \geq 0$ (by strong subadditivity) to deduce
\begin{align}
\sum_a p_a^* I^{(a)}_{n} \geq \sum_{i=0}^{n-1} (\epsilon \mathfrak{p}  \delta_i^{(b)} - 2 \epsilon^2).
\label{eq:Iasanonabeli_cor5}
\end{align} 
Substituting this into (\ref{eq:Iasanonabeli_cor2}) with $i = n$, we obtain
\begin{align}
I^{(b)}_{n+1} \geq \sum_{i=0}^{n-1} (\epsilon \mathfrak{p}  \delta_i^{(b)} - 2 \epsilon^2) - \log |\mathcal{A}|.
\label{eq:Iasanonabeli_cor6}
\end{align} 

Next, we consider (\ref{eq:paIai}) in the special case where $p_a = p_a^*$. Writing out (\ref{eq:paIai}) in this case and rearranging terms, it is not hard to see that it can be rewritten as
$I_i^{(b)} \geq \log\left(\frac{1}{p_b^*}\right) - \delta_i^{(b)}$. Substituting (\ref{eq:Iasanonabeli_cor1}) into this inequality, we deduce that
\begin{align}
I_{n+1}^{(b)} \geq \log\left(\frac{1}{p_b^*}\right) - \delta_i^{(b)}.
\label{eq:Iasanonabeli_cor7}
\end{align}

The last step is to take an appropriate linear combination of (\ref{eq:Iasanonabeli_cor6}) and (\ref{eq:Iasanonabeli_cor7}), such that all the $\delta_i^{(b)}$ terms cancel out. In particular, we multiply
(\ref{eq:Iasanonabeli_cor6}) by $\alpha$ and add it to (\ref{eq:Iasanonabeli_cor7}) multiplied by $(1-\alpha)/n$ and summed over $i = 0,...,n-1$. Here $\alpha$ is a number between $0$ and $1$ which we will determine shortly. The result is
\begin{align*}
I^{(b)}_{n+1} &\geq \alpha \left( \sum_{i=0}^{n-1} (\epsilon \mathfrak{p}  \delta_i^{(b)} - 2 \epsilon^2) - \log |\mathcal{A}| \right) \\
&+ \frac{1-\alpha}{n} \sum_{i=0}^{n-1} \left[\log\left(\frac{1}{p_b^*}\right) - \delta_i^{(b)}\right]. 
\end{align*} 
To make all the $\delta_i^{(b)}$ terms cancel out, we choose $\alpha$ so that $(1-\alpha)/n = \alpha \epsilon \mathfrak{p}$, or equivalently $\alpha = \frac{1}{n \mathfrak{p} \epsilon + 1}$. This gives:
\begin{align*}
I^{(b)}_{n+1} \geq \log\left(\frac{1}{p_b^*}\right) - \alpha \left[2 n \epsilon^2 + \log\left(\frac{|\mathcal{A}|}{p_b^*}\right) \right].
\end{align*}
In order to maximize the right hand side, we choose $\epsilon \sim n^{-1/2}$. More specifically, we take $\epsilon = \frac{\mathfrak{p}}{2\sqrt{n}}$ since this choice ensures that $\epsilon \leq \mathfrak{p}/2$ as required by the inequality (\ref{eq:Iasanonabeli_cor3}), and hence (\ref{eq:Iasanonabeli_cor6}). The result is
\begin{align*}
I^{(b)}_{n+1} \geq \log\left(\frac{1}{p_b^*}\right) - \alpha \left[\frac{\mathfrak{p}^2}{2} + \log\left(\frac{|\mathcal{A}|}{p_b^*}\right) \right].
\end{align*}
Finally, since $\alpha = \frac{1}{n \mathfrak{p} \epsilon + 1} \leq \frac{2}{\mathfrak{p}^2 \sqrt{n}}$, we derive
\begin{align*}
I^{(b)}_{n+1} \geq \log\left(\frac{1}{p_b^*}\right) - \frac{K}{\sqrt{n}}
\end{align*}
where
\begin{align*}
K = 1 + \frac{2}{\mathfrak{p}^2} \log\left(\frac{|\mathcal{A}|}{\mathfrak{p}}\right).
\end{align*}
This completes our proof of (\ref{eq:nonabelineq2}) and hence also (\ref{eq:nonabelineq}).

\subsubsection{Deriving the inequalities (\ref{eq:paIarepeat}) and (\ref{eq:Iasanonabel})}
\label{sec:derivnonabelineq}
All that remains is to derive the inequalities (\ref{eq:paIarepeat}) and (\ref{eq:Iasanonabel}) which formed the basis for the above proof. We start with (\ref{eq:paIarepeat}). This inequality is identical to Eq.~(\ref{eq:paIa}) from the Abelian case, and the derivation is also identical. Again, we consider a mixed state of the form $\lambda = \sum_a p_a \rho^{(a)}$, and using properties \ref{distnonabel}, \ref{indistnonabel}, we deduce that
\begin{align}
I(A:C|B)_\lambda = \sum_a p_a I(A:C|B)_{\rho^{(a)}} - H(\{p_a\})
\label{eq:iabclambdarepeat}
\end{align}
where $H(\{p_a\}) = -\sum_a p_a \log p_a$ is the Shannon entropy. By strong subadditivity, we derive (\ref{eq:paIarepeat}).

To derive (\ref{eq:Iasanonabel}), consider the mixed state $\lambda = \sum_a p_a \rho^{(a)}$. Letting $W_s$ be the unitary operator from property \ref{fusionnonabel} above, notice that
\begin{align}
I(A:C|B)_{\lambda} &= I(A:C|B)_{W_s \lambda W_s^\dagger} \nonumber \\
&\geq  I(A':C|B)_{W_s \lambda W_s^\dagger}. 
\label{eq:iabcnonabel}
\end{align}
Here the first line follows from the fact that $W_s$ is a unitary operator supported in region $A$ and therefore does not affect the conditional mutual information. The second line follows from the general fact that $I(A:C|B)$ is non-increasing under tracing out spins in $A$ or $C$ (a corollary of strong subadditivity). Next, using (\ref{eq:fusionassumpnonabel}), we write the density operator $(W_s \lambda W_s^\dagger)_{A' B C}$ as a mixture of the form
\begin{align}
(W_s \lambda W_s^\dagger)_{A' B C} = \sum_a p_{a,s} \rho^{(a)}_{A'BC}
\label{eq:wsmix}
\end{align}
where $p_{a,s}$ is defined as in (\ref{eq:pas}). Finally, we note that (\ref{eq:wsmix}) means that $I(A':C|B)_{W_s \lambda W_s^\dagger}$ can be expanded out similarly to (\ref{eq:iabclambdarepeat}):
\begin{align}
I(A':C|B)_{W_s \lambda W_s^\dagger} = \sum_a p_{a,s} I(A':C|B)_{\rho^{(a)}} - H(\{p_{a,s}\})
\label{eq:iabcv2}
\end{align}
Substituting (\ref{eq:iabcv2}) and (\ref{eq:iabclambdarepeat}) into (\ref{eq:iabcnonabel}), gives the desired inequality (\ref{eq:Iasanonabel}).

\begin{figure}[tb]
   \centering
   \includegraphics[width=0.5\columnwidth]{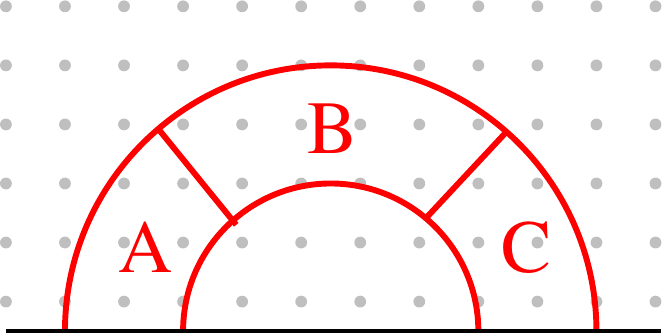} 
   \caption{Geometry used to compute the boundary topological entanglement entropy $\gamma_{\text{bd}}$.}
   \label{fig:TEE_boundary}
\end{figure}

\section{Discussion}
\label{sec:discuss}

In this paper, we have presented a proof of the two inequalities (\ref{eq:main_result}) and (\ref{eq:main_result_v2}) starting from a small set of physical assumptions about $\rho$. We emphasize that our proof, like that of Ref.~\onlinecite{teelowerbound}, relies heavily on a specific definition of the TEE -- namely the definition in Eq.~(\ref{eq:TEEdef}), originally given in Ref.~\onlinecite{levinwen2006}. We do not know whether these inequalities hold for the definition of TEE given in Ref.~\onlinecite{kitaevpreskill2006}, or for the cylinder extrapolation definition discussed in Ref.~\onlinecite{Zou_2016}.

Due to the generality of our arguments, our proof can be extended to a variety of systems. The simplest extension is to \emph{fermionic} states. This extension comes for free since our derivation only uses the strong subadditivity property of the von Neumann entropy, which is known to hold for both fermionic~\cite{araki2003equilibrium} and bosonic systems~\cite{Lieb1973}. The only subtlety in the fermionic case is the definition of the anyon types $\mathcal{A}$: to make our derivation of (\ref{eq:main_result}) and (\ref{eq:main_result_v2}) go through in the fermionic case, we need to define $\mathcal{A}$ to be the set of anyon types \emph{modulo} the local fermion. 

Another straightforward extension is to systems with gapped boundaries. For such systems, one can define a ``boundary topological entanglement entropy'' -- an analog of the TEE that probes the properties of boundary excitations~\cite{shi2021domain, shi2021entanglement}. The boundary TEE is defined by $\gamma_{\text{bd}} = \frac{1}{2} I(A:C|B)_\rho$ where $\rho$ is the density operator for the ground state and $A, B, C$ are non-overlapping subsets of the lattice arranged in a half-annulus geometry as in Fig.~\ref{fig:TEE_boundary}. Again, our assumptions and arguments carry over to this situation, leading to the same two inequalities (\ref{eq:main_result}) and (\ref{eq:main_result_v2}), but with $\gamma$ replaced by $\gamma_{\text{bd}}$. The only change is that $\mathcal{A}$ now corresponds to the set of \emph{boundary} anyons, and $\mathcal{D}$ is the total quantum dimension of the boundary, i.e.~$\mathcal{D} = \sqrt{\sum_a d_a^2}$ where the sum runs over $a \in \mathcal{A}$.

A slightly more nontrivial extension is to systems with point-like defects, e.g.~the model discussed in Refs.~\onlinecite{bombin2010, browntwist2013}. The main complication in these systems is that the set of anyon sectors near the defect is not necessarily in one-to-one correspondence with the set of mobile anyon excitations, or equivalently, the set of flexible string operators. As a result, we need to modify our assumptions if we wish to describe a geometry with a defect inside the annulus $ABC$. The first modification is to label the string operators $W_s$ from property \ref{fusionnonabel} using a different set $s \in \mathcal{S}$ than the set $\mathcal{A}$ that labels the anyon sectors. (For example, in the example of Ref.~\onlinecite{bombin2010, browntwist2013}, the string operators would be labeled as $\mathcal{S} = \{1,e,m, \epsilon\}$ while the anyon sectors would be labeled as $\mathcal{A} = \{\sigma_+, \sigma_-\}$).  Likewise, when we define the fusion probabilities $p_{s \times a \rightarrow b}$, we need to take the index $s \in \mathcal{S}$, while $a, b \in \mathcal{A}$. Finally, we need to modify our assumption (\ref{eq:probid}) about the fixed-point probability distribution $p_a^*$. In particular, we need to allow for the possibility of two different distributions -- a distribution $q_s^*$ defined on $s \in \mathcal{S}$, and a distribution $p_a^*$ defined on $a \in \mathcal{A}$, which together obey the analog of (\ref{eq:probid}), namely $\sum_s  p_{s \times b \rightarrow a} q_s^* = p_a^*$. Once we modify our assumptions in this way, the proof proceeds exactly as before, and we deduce that the inequality (\ref{eq:nonabelineq}) holds even in the presence of the defects.

Our results can also be extended to three dimensional (3D) systems. In 3D systems, we can define two quantities analogous to the TEE~\cite{castelnovo3dtee2008}. The first quantity is defined by $\gamma_{\text{sph}} = \frac{1}{2} I(A:C|B)_\rho$, where $A, B, C$ are non-overlapping regions arranged in a spherical shell geometry with $A$ and $C$ being caplike regions on the two ends of the sphere and $B$ spanning the remainder of the spherical shell. (This geometry is equivalent to the one shown in Fig.~\ref{fig:TEE_schematic} but rotated about the horizontal axis). The second quantity is defined by $\gamma_{\text{tor}} = \frac{1}{2} I(A:C|B)_\rho$, where $A, B, C$ are non-overlapping regions arranged in a solid torus geometry with $A$ and $C$ being solid cylinders on the two ends of the solid torus and $B$ spanning the remainder of the solid torus. (This geometry is equivalent to the one shown in Fig.~\ref{fig:TEE_schematic} but thickened in the direction perpendicular to the plane). In both cases, we can apply the same arguments as before and derive the two inequalities (\ref{eq:main_result}) and (\ref{eq:main_result_v2}), but with $\gamma$ replaced by $\gamma_{\text{sph}}$ or $\gamma_{\text{tor}}$ respectively. In the first case,  $\mathcal{A}$ corresponds to the set of particle-like excitations, while in the second case $\mathcal{A}$ corresponds to the set of loop-like excitations. In both cases, $\mathcal{D}$ is defined by $\mathcal{D} = \sqrt{\sum_a d_a^2}$ where the sum runs over $a \in \mathcal{A}$. One caveat is that, in this discussion, we implicitly assumed that all the excitations in our system -- whether particle-like or loop-like -- are mobile. This assumption breaks down in fracton systems so more care may be needed to address these systems~\cite{ma2018topological}. 

Finally, we note that our results also apply if $\rho$ is a \emph{mixed} state. The only subtlety in the mixed state case is that we may want to use a slightly weaker version of assumption \ref{fusionnonabel}: instead of assuming the existence of unitary operators $W_s$ obeying (\ref{eq:fusionassumpnonabel}), it may be more natural to assume the existence of quantum channels $\mathcal{W}_s$ obeying the analog of (\ref{eq:fusionassumpnonabel}) with $W_s \rho^{(a)} W_s^\dagger$ replaced with $\mathcal{W}_s[\rho^{(a)}]$. Conveniently, our derivation goes through exactly as before, even with this weaker assumption. Recently, there has been growing interest in finding diagnostics for identifying mixed-state topological phases~\cite{fan2024diagnostics}. Our results suggest that the TEE, as defined in Eq.~(\ref{eq:TEEdef}), may have promise as such a diagnostic.

\acknowledgments

We would like to thank Isaac Kim, Ting-Chun Lin, Daniel Ranard, Bowen Shi and Chenjie Wang for discussions. M.L. acknowledges the support of the Kadanoff Center for Theoretical Physics at the University of Chicago. This work was supported in part by a Simons Investigator grant and by the Simons Collaboration on Ultra-Quantum Matter, which is a grant from the Simons Foundation (651442).

\appendix

\section{Deriving (\ref{eq:probid}) from the algebraic theory of anyons}
\label{sec:probidapp}
In this Appendix, we show that, for any system described by the algebraic theory of anyons~\cite{Kitaev_2006}, the probability distribution $p_a^* = \frac{d_a^2}{\mathcal{D}^2}$ obeys the identity (\ref{eq:probid}). That is,
\begin{align}
\sum_s p_{s \times a \rightarrow b} \frac{d_s^2}{\mathcal{D}^2} = \frac{d_b^2}{\mathcal{D}^2}
\label{eq:probidrepeat}
\end{align}
for all $a, b \in \mathcal{A}$.

The first step is to recall that according to the standard anyon formalism, the fusion probabilities $p_{s \times a \rightarrow b}$ are given by~\cite{preskilltqc, kitaevpreskill2006}
\begin{align}
p_{s \times a \rightarrow b} = \frac{d_b N^{sa}_b}{d_s d_a}
\end{align}
where $N^{ab}_c$ is the fusion multiplicity defined by $a \times b = \sum_c N^{ab}_c c$. 

We can then derive (\ref{eq:probidrepeat}) as follows:
\begin{align}
\sum_s p_{s \times a \rightarrow b} \frac{d_s^2}{\mathcal{D}^2} &= \sum_s \frac{d_b d_s N^{sa}_b}{d_a \mathcal{D}^2} \nonumber \\
&= \sum_s \frac{d_b d_s N^{b \bar{a}}_{s}}{d_a \mathcal{D}^2} \nonumber \\
&= \frac{d_b^2}{\mathcal{D}^2} 
\end{align}
where the second line follows from the property $N^{sa}_b =  N^{b \bar{a}}_{s}$, while the last line follows from the standard identity $\sum_c N^{ab}_c d_c = d_a d_b$, together with the fact that $d_a = d_{\bar{a}}$~\cite{Kitaev_2006}.

\section{Deriving (\ref{eq:probid}) from two conditions on fusion probabilities}
\label{sec:probidapp2}
In this Appendix, we derive the existence of a fixed point probability distribution $p_a^*$ obeying (\ref{eq:probid}), from two conditions on the fusion probabilities $p_{s \times a \rightarrow b}$. Specifically, the two conditions are: (i) for every $a, b \in \mathcal{A}$, there exists an $s \in \mathcal{A}$ with $p_{s \times a \rightarrow b} > 0$ and (ii) the fusion probabilities are associative in the sense that
\begin{align}
\sum_u p_{s \times t \rightarrow u} p_{u \times a \rightarrow c} = \sum_b  p_{t \times a \rightarrow b} p_{s \times b \rightarrow c}.
\label{eq:assoc_cond}
\end{align}

We begin with some notation. Let $p_s$ and $q_a$ be two probability distributions. We define $p \ast q$ to be the probability distribution
\begin{align}
(p \ast q)_b = \sum_{sa}  p_{s \times a \rightarrow b} p_s q_a.
\end{align}
That is, $(p \ast q)_b$ is the probability distribution obtained by picking an anyon $s$ with probability $p_s$ and an anyon $a$ with probability $q_a$, and then fusing them together. An important property of this ``$\ast$'' operation is that it is associative:
\begin{align}
p \ast (q \ast r) = (p \ast q) \ast r
\label{eq:assoc_cond2}
\end{align}
for any three probability distributions $p, q, r$. Indeed, this associativity property is equivalent to (\ref{eq:assoc_cond}) above.

Let $p$ be the uniform distribution, $p_s = 1/|\mathcal{A}|$, and consider the matrix $M_{ab}$ defined by $M_{ab} = \sum_s  p_{s \times a \rightarrow b} p_s$. By condition (i) above, all the matrix elements $M_{ab}$ are strictly positive: $M_{ab} > 0$. Therefore, by the Perron-Frobenius theorem, there exists a unique probability distribution $q_a$ such that
\begin{align}
\sum_a M_{ab} q_a = q_b.
\end{align}
Rewriting this equation using the ``$\ast$'' notation, we deduce that there exists a unique probability distribution $q$ such that
\begin{align}
p \ast q = q.
\label{eq:fixedpteq}
\end{align}

We claim that $p_s^* \equiv q_s$ obeys (\ref{eq:probid}) -- that is $q$ is the desired fixed-point distribution. To see this, consider any other probability distribution $r$. By (\ref{eq:assoc_cond2}), we have
\begin{align}
p \ast (q \ast r) &= (p \ast q) \ast r \nonumber \\
&= q \ast r.
\end{align}
Then by the uniqueness of $q$ as a solution to (\ref{eq:fixedpteq}), we deduce that
\begin{align}
q \ast r = q.
\end{align}
Letting $r$ be the probability distribution $r_a = \delta_{ab}$ and plugging in the definition of $\ast$, this equation reduces to (\ref{eq:probid}), as we wished to show.

\section{Derivation of (\ref{eq:Iasanonabeli_cor3})}
\label{sec:nonabelcorapp}
In this Appendix, we derive the inequality (\ref{eq:Iasanonabeli_cor3}), which we reprint below for convenience:
\begin{align}
\sum_a p_a^* (I^{(a)}_{i+1} -  I^{(a)}_i) &\geq \epsilon \mathfrak{p} \left[I^{(c)}_{i} - I^{(b)}_i + \log\left( \frac{p_{c}^*}{p_b^*}\right) \right] - 2 \epsilon^2.
\label{eq:Iasanonabeli_cor3_repeat}
\end{align} 
We will establish this inequality for any pair of anyons $b, c \in \mathcal{A}$ and any $\epsilon$ with $|\epsilon| \leq \mathfrak{p}/2$, where $\mathfrak{p} = \min_a p_a^*$.

The first step is to consider the inequality (\ref{eq:Iasanonabeli2}) for a probability distribution of the form 
\begin{align}
p_a = p_a^* + \epsilon \delta_{a b} - \epsilon \delta_{a c}.
\label{eq:probdistep}
\end{align}
Specializing to such probability distributions, the inequality (\ref{eq:Iasanonabeli2}) simplifies to
\begin{align}
\sum_a p_a (I^{(a)}_{i+1} -  I^{(a)}_i) \geq \epsilon  &(I^{(c)}_{i} - I^{(b)}_i)  \nonumber \\
&+ H(\{p_a\}) - \sum_s p_s^* H(\{p_{a,s}\}). 
\label{eq:Iasanonabeli3}
\end{align}
The next step is to find a lower bound for the difference $H(\{p_a\}) - \sum_s p_s^* H(\{p_{a,s}\})$ using a first order Taylor series expansion (in $\epsilon$). Below we will show that such a Taylor series expansion gives the following lower bound which is valid for any $|\epsilon| \leq  \mathfrak{p}/2$:
\begin{align}
H(\{p_a\}) - \sum_s p_s^* H(\{p_{a,s}\}) \geq  \epsilon \log \left(\frac{p_{c}^*}{p_b^*} \right) - \frac{2 \epsilon^2}{\mathfrak{p}}.
\label{eq:Hpaineq}
\end{align}
We postpone the derivation of (\ref{eq:Hpaineq}) to the end of this Appendix. Substituting (\ref{eq:Hpaineq}) into (\ref{eq:Iasanonabeli3}), we obtain the inequality
\begin{align}
\sum_a p_a (I^{(a)}_{i+1} -  I^{(a)}_i) \geq \epsilon  \left[I^{(c)}_{i} - I^{(b)}_i + \log\left( \frac{p_{c}^*}{p_b^*}\right) \right] - \frac{2 \epsilon^2}{\mathfrak{p}}.
\label{eq:Iasanonabeli4}
\end{align}
At this point, we have almost derived (\ref{eq:Iasanonabeli_cor3_repeat}), except that the left hand side is written in terms of $p_a$ instead of $p_a^*$. To complete the derivation, we note that
\begin{align}
\sum_a p_a (I_{i+1}^{(a)} - I_i^{(a)}) &\leq \max_a (I_{i+1}^{(a)} - I_i^{(a)}) \nonumber \\
&\leq \frac{1}{\mathfrak{p}} \sum_a p_a^* (I_{i+1}^{(a)} - I_i^{(a)})
\label{eq:papastarineq}
\end{align}
where the second inequality follows from the fact that $p_a^* \geq \mathfrak{p}$ for all $a$, together with the fact that $I_{i+1}^{(a)} - I_i^{(a)} \geq 0$ by
(\ref{eq:Iasanonabeli_cor1}). Combining (\ref{eq:Iasanonabeli4}) with (\ref{eq:papastarineq}) gives the desired inequality (\ref{eq:Iasanonabeli_cor3_repeat}).

All that remains is to derive the lower bound (\ref{eq:Hpaineq}). As we mentioned earlier, we do this by considering the first order Taylor series expansion for $H(\{p_a\}) - \sum_s p_s^* H(\{p_{a,s}\})$. First, recall the statement of the (first order) Taylor's theorem. Let $f(x)$ be a function that is twice differentiable and has a continuous second derivative on an open interval containing $x_0$ and $x$. Then, Taylor's theorem states that
\begin{align}
f(x) = f(x_0) + f'(x_0)(x-x_0) + \frac{f''(y)}{2} (x-x_0)^2
\label{eq:taylor}
\end{align}
for some $y$ between $x_0$ and $x$. Applying this theorem to $f(x) = -x \log x$ with $x_0 = p_a^*$ and $x = p_a$ we obtain
\begin{align}
-p_a \log p_a = -p_a^* \log p_a^* - (\log p_a^* &+ 1) (p_a-p_a^*) \nonumber \\
&- \frac{1}{2 y} (p_a-p_a^*)^2
\end{align}
for some $y$ between $p_a$ and $p_a^*$. To proceed further, notice that $p_a \geq \mathfrak{p}/2$ for all $a \in \mathcal{A}$ since $|\epsilon| \leq  \mathfrak{p}/2$ and $\mathfrak{p} = \min_a p_a^*$. Hence, we must also have $y \geq  \mathfrak{p}/2$. We therefore have the lower bound
\begin{align}
-p_a \log p_a \geq -p_a^* \log p_a^* - (\log p_a^* &+ 1) (p_a-p_a^*) \nonumber \\
&- \frac{1}{\mathfrak{p}} (p_a-p_a^*)^2.
\end{align} 
Summing over all $a$ and using the specific form of our probability distribution $p_a$ (\ref{eq:probdistep}) gives 
\begin{align}
H(\{p_a\}) \geq H(\{p_a^*\}) + \epsilon \log \left(\frac{p_c^*}{p_b^*} \right) - \frac{2  \epsilon^2}{\mathfrak{p}}.
\label{eq:lowerboundH}
\end{align}

So far we have derived a lower bound on $H(\{p_a\})$. To complete our derivation of (\ref{eq:Hpaineq}), we also need an \emph{upper} bound on $\sum_s p_s^* H(\{p_{a,s}\})$. For that, we use the fact that $H$ is a concave function. This concavity property implies that 
\begin{align}
\sum_s p_s^* H(\{p_{a,s}\}) \leq H(\{p_a^*\}) 
\label{eq:upperboundH}
\end{align}
since $\sum_s p_s^* p_{a,s} = p_a^*$. Combining (\ref{eq:lowerboundH}) with (\ref{eq:upperboundH}), proves the claim (\ref{eq:Hpaineq}).


\end{document}